\pdfoutput=1

\documentclass{aa}  

\usepackage{graphicx}
\usepackage{natbib}
\usepackage{xcolor}

\usepackage{subfig}
\usepackage{txfonts}
\usepackage[]{hyperref}

\usepackage{silence}
\newcommand{\xmm}{XMM-\emph{Newton}}

\newcommand{\nustar}{\textit{NuSTAR}}
\newcommand{\hst}{{\it HST}}
\newcommand{\ms}{M$_{\odot}$}

\newcommand{\ls}{L$_{\odot}$}
\newcommand{\kms}{km s$^{-1}$}
\newcommand{\fluxcgs}{ergs s$^{-1}$ cm$^{-2}$}
\newcommand{\lumcgs}{ergs s$^{-1}$}

\newcommand{\aco}{$\alpha_{\rm CO}$}
\newcommand{\barolo}{$^{\rm 3D}$BAROLO}
\newcommand{\uaco}{M$_{\odot}$ pc$^{-2}$ (K km s$^{-1}$)$^{-1}$}
\newcommand{\uLco}{K km s$^{-1}$ pc$^2$}
\begin{document} 

\title{The molecular gas in the central region of NGC 7213}
\author{F. Salvestrini\inst{\ref{inst1},\ref{inst2}}
\and
C. Gruppioni\inst{\ref{inst2}}
\and
F. Pozzi\inst{\ref{inst1},\ref{inst2}}
\and
C. Vignali\inst{\ref{inst1},\ref{inst2}}
\and 
A. Giannetti\inst{\ref{inst3}}
\and 
R. Paladino\inst{\ref{inst3}}
\and
E. Hatziminaoglou\inst{\ref{inst4}}}

\institute{
Dipartimento di Astronomia, Universit\`a degli Studi di Bologna, Via Gobetti 93/2, 40129 Bologna, Italy\label{inst1}
\and
INAF - Osservatorio di Astrofisica e Scienza dello Spazio di Bologna, Via Gobetti 93/3 - 40129 Bologna - Italy\label{inst2}
\and
INAF - Istituto di Radioastronomia \& Italian ALMA Regional Centre, Via P. Gobetti 101, 40129 Bologna, Italy\label{inst3}
\and
ESO, Karl-Schwarzschild-Str 2, D-85748 Garching bei M\"unchen, Germany\label{inst4}}

%
%
\abstract{
We present a multi-wavelength study (from X-ray to millimetre) of the nearby low-luminosity active galactic nucleus (LLAGN) NGC 7213.
We combine the information from the different bands to characterise the source in terms of contribution from the AGN and the host-galaxy interstellar medium (ISM).
This approach allows us to provide a coherent picture of the role of the AGN and its impact, if any, on the star formation and molecular gas properties of the host galaxy.\\
We focused our study on archival ALMA Cycle 1 observations, where the CO(2-1) emission line has been used as a tracer of the molecular gas. 
Using the \barolo\ code on ALMA data, we performed the modelling of the molecular gas kinematics traced by the CO(2-1) emission, finding a rotationally dominated pattern.
The host-galaxy molecular gas mass was estimated from the integrated CO(2-1) emission line obtained with APEX data, assuming an \aco\ conversion factor.
By using the ALMA data, we would have underestimated the gas masses by a factor $\sim$3, given the filtering out of the large scale emission in interferometric observations.
We also performed a complete X-ray spectral analysis on archival observations, revealing a relatively faint and unobscured AGN.
The AGN results to be too faint to significantly affect the properties of the host-galaxy, such as star formation activity and molecular gas kinematics and distribution.}

 \maketitle
%
%
\section{Introduction}
Active galactic nuclei (AGN) are thought to play a key role in regulating the host-galaxy star formation (SF).
The accretion of matter onto the central supermassive black hole (SMBH) is responsible for injecting energy in the circum-nuclear region, providing feedback to its host galaxy and the interstellar medium (ISM) (see, e.g., \citealt{Fabian12}; \citealt{SomervilleDave15}, and references therein). 
For this reason, the SF activity and SMBH properties are believed to be connected, both in high-redshift quasars and in local Seyfert nuclei.
AGN are held responsible for both suppressing the star formation rate, SFR, (i.e. {\it negative feedback}) or enhancing it through the compression of molecular clouds (i.e. {\it positive feedback}).\\
In this scenario, the molecular gas plays a fundamental role, being the main fuel for SF and the more abundant phase of the ISM in the nuclear region.
Therefore, studying the properties of the molecular gas in galaxies and the rate at which it is converted into stars (depletion time, t$_{\rm depl}=$M$_{gas}$/SFR) is crucial to understand the processes at play in galaxies. 
If the AGN is able to completely remove or heat the gas, thus preventing it from cooling, we would expect low t$_{\rm depl}$ values with respect to inactive galaxies with similar stellar masses (M$_{\star}$) and SFR. 
Indeed, t$_{\rm depl}$ (0.01$<$t$_{\rm depl}<$0.1 Gyrs) lower than in normal galaxies with similar SFR and M$_{\star}$ have been found in luminous AGN at high redshift (i.e. z$\sim$1.5-2.5, \citealt{Kakkad17}; \citealt{Brusa18}; \citealt{Talia18}), while in the local Universe this effect is not completely understood (e.g., \citealt{GarciaBurillo14}; \citealt{Casasola15}; \citealt{Rosario18}).\\
A multi-wavelength approach is necessary to fully characterize the mechanisms regulating the relation between the accretion onto the SMBH and the SF process within its host galaxy.
Spatially and spectrally resolved observations, tracing the cold phase of the ISM, are necessary to understand the impact of the AGN.
In fact, if the AGN is present, it can dominate the emission close to the nuclear regions, while at increasing distances from the centre stellar processes such as supernovae, stellar winds or shocks start dominating.
The high spatial resolution and high sensitivity provided by the Atacama Large Millimeter Array (ALMA) are crucial to study the feeding and feedback processes that could take place at sub-kpc scales, near the nucleus.
Coupling this information with single-dish observation, needed to recover the whole content of molecular gas in galaxies, allows us to characterise the properties of the molecular component.
This, in combination with the modelling of the spectral energy distribution (SED), using broad-band photometry from the UV/optical to the far infrared (FIR), can allow us to constrain the contributions of stellar processes and the AGN to the global output of the source. 
Eventually, X-ray observations, especially in the hard band, directly probe the accretion-related emission from the nuclear region, hence the radiating power of the AGN.
To obtain a complete picture of the interplay between the AGN and the host-galaxy, it is necessary to combine in a coherent way all the information from the different wavebands. 
\\
\cite{Gruppioni16} (G16, hereafter) presented the results of a detailed broad-band SED decomposition on a statistically significant sample of local active galaxies, including the emission from stars, dust-reprocessed emission by SF and AGN dusty torus.
The sample consisted of 76 nearby active galaxies (i.e. 36 Seyfert 1, 37 Seyfert 2, and 3 low-ionisation narrow emission-line region, LINER) from the complete sample of active and in-active galaxies in the local Universe, selected at 12 $\mu$m by \cite{RMS93}.
In particular, the 76 sources presented in G16 were selected from the parent sample among the active galaxies based on the availability of a \emph{Spitzer}/IRS spectra in the MIR.
Combining the MIR information with an ancillary collection of photometric data from the optical to the FIR, the analysis of the broad-band SED allowed us to derive M$_\star$, SFR, M$_{\rm dust}$, and the IR luminosity from the AGN and from the SF.
To assess whether and to what extent the AGN is able to regulate the host-galaxy SF, we need to study the properties of the molecular gas, which is the main fuel of the SF activity.
This goal can be achieved by combining the information on the morphology and kinematics of the molecular gas obtained with high-resolution observations in the millimeter band, with the determination of the relative contribution of the AGN to the global outcome of the galaxy (obtained using the SED decomposition and the characterisation of the AGN power through the analysis of the emission in the X-rays).\\
In this work, we present a test study to show the potential of this multi-waveband method, focusing our attention on one object out of the 76 by G16.
The target of this study is NGC 7213, a nearby spiral galaxy showing intermediate properties between a low-luminosity AGN (LLAGN) and a LINER.
The source was chosen due to the quality of the available archival observations in different bands, in particular in the X-rays (e.g., \nustar\ and \xmm) to characterise the AGN power, and at mm wavelengths (ALMA and APEX) to trace the molecular gas content and kinematics.\\
The paper is organised as follows: in Sect. \ref{sec:ngc7213} we summarise the multi-waveband properties of the NGC 7213.
In Sect. \ref{sec:intro_data} we introduce the data sets that have been reduced and analysed in this work.
The interpretation of the CO and the continuum mm emission is presented in Sect. \ref{sec:interpretation}.
The conclusions and the results are summarised in Sect. \ref{sec:conclusions}.\\
%
%
\section{NGC 7213}
\label{sec:ngc7213}
NGC 7213 is a nearby (D=23 Mpc, z=0.0058) S0 galaxy, hosting an active nucleus, first discovered with the \emph{HEAO} A-2 satellite (e.g., \citealt{Marshall79}).
The classification of this source was long debated (e.g., \citealt{HalpernFilippenko84}), and nowadays it is known as an intermediate object --- between a LLAGN (with L$_{\rm bol}=1.7\times10^{43}$ \lumcgs, \citealt{Emmanoulopoulos12}) and a LINER (e.g., \citealt{Starling05_I}).
The first published optical spectrum by \cite{Phillips79} suggested the Seyfert 1 classification on the basis of the observed broad H$\alpha$ emission line component (with full width at zero intensity $\sim$13000 \kms).
They also found that the flux of the H$\alpha$ was relatively low with respect to what usually measured in typical Seyfert 1 galaxies, and broad components were very weak or absent in the other observed optical emission lines.
Later, \cite{HalpernFilippenko84} confirmed the presence of the broad H$\alpha$ emission line, but the evidence for a low-excitation narrow-line spectrum led to the inclusion of the source in the LINER class.\\
The X-ray observations confirmed the ambiguous nature of NGC 7213.
Archival observations with different X-ray telescopes over several years showed some spectral features in agreement with the Seyfert 1 classification (e.g., an X-ray spectral slope $\Gamma_{X}\sim1.8$ and no evidence for neutral or ionised absorption features; \citealt{Bianchi08}; \citealt{Lobban10}; \citealt{Emmanoulopoulos13}), while others did not (e.g., the absence of a Compton reflection component, usually observed in local Seyfert 1 galaxies; e.g., \citealt{Dadina08}; \citealt{Ursini15}).\\
To complete the multi-band picture of NGC 7213, at radio frequencies the galaxy appears point-like at 3 cm (half power beam width HPBW$\lesssim$1 arcsec), which was interpreted by \cite{Bransford98} as either nuclear synchrotron emission or free-free emission.
The radio power is P$_{\rm 1.4 GHz}=3\times10^{29}$ erg s$^{-1}$ Hz$^{-1}$, at least an order of magnitude higher than that of a typical Seyfert, although too low for a radio-loud classification (e.g., \citealt{BlankHarnett05}).
The compactness of the radio emission has been later confirmed also by \cite{Murphy10} who observed NGC 7213 with ATCA at higher frequencies (5, 8, and 20 GHz; see also \citealt{Bell11}).
%
%
\section{The multi-wavebands data}
\label{sec:intro_data}
In this work, we complete the multi-band picture of NGC 7213 by providing a new and coherent modelling of the most relevant data to describe the overall emission of the source in the X-rays over a broad energy range (including the hard-X data from \nustar) and the analysis of sub-mm/mm single-dish (APEX) and interferometric (ALMA) observations.
The X-ray data analysis aims at providing an accurate estimate of the accretion power, while the study of the high spatial resolution of the ALMA  data is used to characterise the morphology and kinematics of the molecular gas.
Finally, the single-dish APEX observation is used to provide a reliable estimates of the integrated CO emission, needed to derive the molecular gas mass content. 
\subsection{X-ray data}
NGC 7213 has been observed extensively in the last 20 years in the X-rays using a number of facilities, in both soft and hard bands.
We are interested in characterising the nuclear activity of the source in terms of the emitting power of the AGN, i.e. the luminosity in the 2-10 keV band produced by the primary emission.
For this reason, we decided to use the largest band available, combining the information from \nustar\ (nominally, 3-79 keV) with an instrument in the 0.3-10 keV band (i.e. \xmm).
We analysed separately the \nustar\ observation and the one from \xmm\ with the longer exposure time (130 ks; e.g., \citealt{Emmanoulopoulos13}) to obtain a global picture of the properties of the source in terms of the spectral features and continuum emission.
We did not combine the \nustar\ and \xmm\ observations since they were not taken simultaneously and previous works revealed evidences for minor variability in terms of flux and spectral features in NGC 7213 (e.g., \citealt{Ursini15}). 
Nevertheless, the observed variability do not significantly affect the X-ray properties of the source (e.g., \citealt{Bianchi03}; \citealt{Lobban10}; \citealt{Emmanoulopoulos13}).
%
%
\subsubsection{X-ray data reduction}
In this work we re-analyse and combine the following archival observations from \xmm\  (ID: 605800301; starting in Nov. 11th 2009; t$_{\rm exp}=132.5$ ks) and \nustar\ (ID: 60001031002; starting in Oct. 05th 2014; t$_{\rm exp}=101.6$ ks). 
We performed a standard data reduction for each dataset, using the dedicated softwares: the Science Analysis Software (\texttt{SAS}) v.16.1.0 for \xmm, the \texttt{HEASOFT} (v. 6.19) distribution for \nustar\  Focal Plane Modules (FPM; \texttt{NuSTARDAS}, NuSTAR Data Analysis Software v1.7.1).
During data reduction, we checked the light curves for potential time variability, once flaring-background periods were filtered out.
No evidence for significant time variability during the observations was found.
For each observation, the source counts were extracted from circular regions centered on the radio position of NGC 7213, provided by the NASA/IPAC Extragalactic Database (NED)\footnote{https://ned.ipac.caltech.edu/}.
The adopted apertures were chosen depending on the encircled energy fraction (EEF): 15$^{\prime\prime}$ for \xmm\ (corresponding to the 90\% of the EEF below 5 keV for both EPIC pn and MOS cameras), and 60$^{\prime\prime}$ for \nustar\ (corresponding to the 50\% of the EEF on the entire band for both FPM cameras).
The corresponding background counts were extracted from regions free from sources, close to the target, with circular apertures similar to those used for the source.\\
We excluded the energy channels where either calibration issues are known to affect the cameras response, or high-background was present (i.e. signal-to-noise ratio $\sim$1).
In particular, for the \xmm\ pn and MOS cameras, we excluded the channels corresponding to energies below 0.5 keV due to calibration issues, while above 10 keV the background dominates.
Furthermore, the soft band below 2 keV is dominated by thermal emission, associated with the high-energy tail of SF-related emission (e.g., \citealt{Bianchi03}; \citealt{Starling05_I}; \citealt{Bianchi08}; \citealt{Lobban10}). 
Since we are not interested in interpreting this emission, we excluded the channels below 2 keV from our analysis.
Regarding \nustar, as high background affected all the channels above 27 keV, while calibration issues affected the channels below 3 keV, these energy intervals were also removed.
Both \nustar\ and \xmm\ data were grouped with a minimum number of 30 counts in each channel bin.
\begin{figure}[t]
	\includegraphics[width = 0.5\textwidth, keepaspectratio=True]{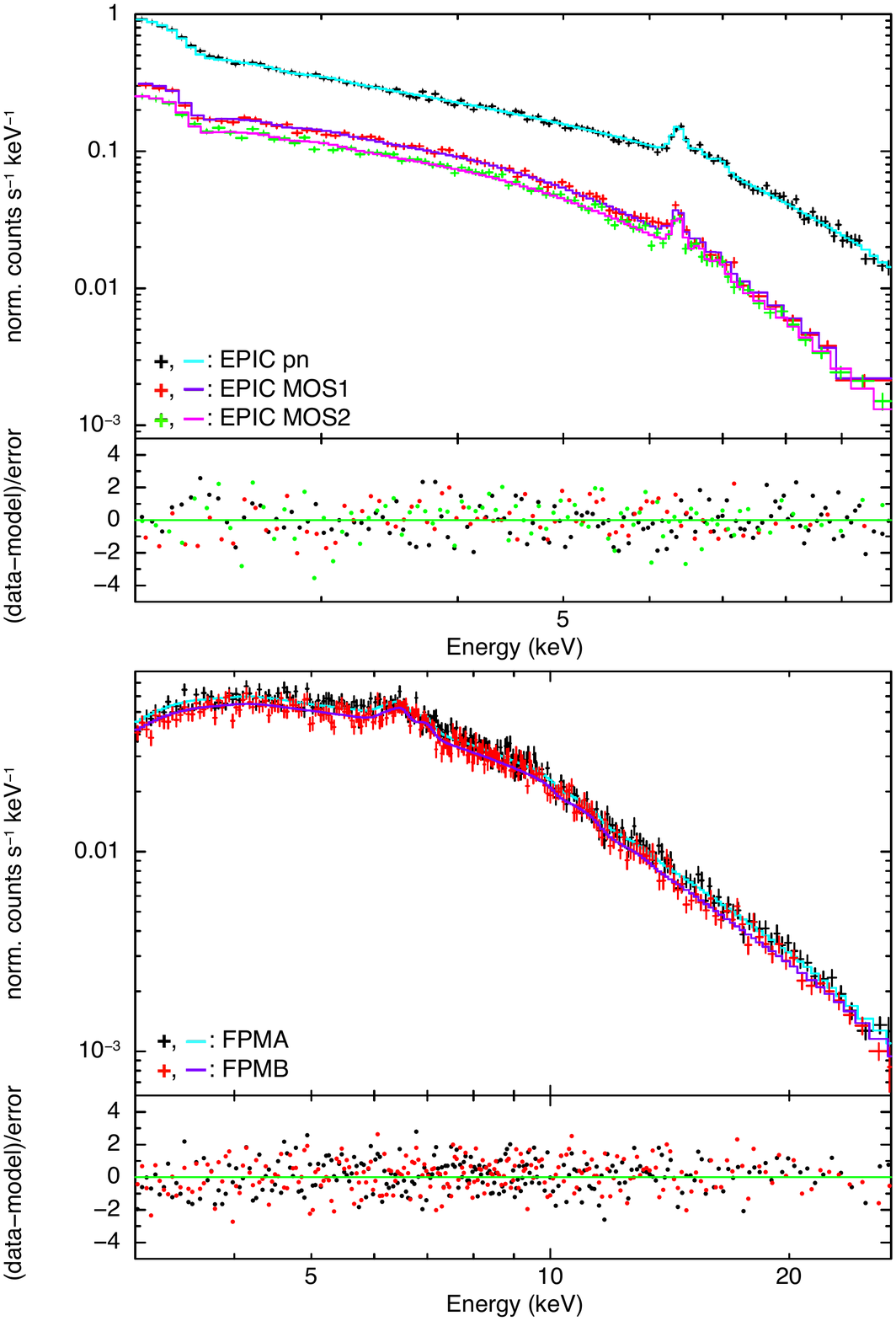}
	\caption{\emph{From top to bottom, the X-ray spectrum of NGC 7213 obtained with \xmm\ (EPIC pn, MOS1 and MOS2) and \nustar\ (FPMA and FPMB), as a function of the observed-frame energies.
	Data and the best-fit models for each camera are represented in different colours.
	In both lower panels, we present the residuals (data minus model) in units of $\sigma$.}}
	\label{fig:xrays}
\end{figure}
%
%
\subsubsection{X-ray data analysis}
\label{sec:xray-analysis}
The spectral analysis has been performed using the X-Ray Spectral Fitting Package (\texttt{XSPEC}) v. 12.10.0c \citep{xspec}.
All the models presented below include the Galactic absorption (N$_{\rm H}=1.06\times10^{20}$ cm$^{-2}$; \citealt{Kalberla05}).
We also included cross calibration constants to account for different response between EPIC pn and both MOS cameras in \xmm, and between FPM A and B in \nustar.
We analysed separately each data set and compared our best-fit models with the literature (e.g., \citealt{Bianchi03} for \xmm, and \citealt{Ursini15} for \nustar).
This allows us also to check for any potential variability both in flux and in spectral shape as a function of time, since the observations were taken with a separation of 5 years.\\
Starting from the \xmm\ observation, the simplest model we used was a single power-law, with best-fit spectral index $\Gamma_{X}\sim1.65$.
This represents the primary X-ray emission from the nuclear activity, produced by inverse-Compton of the hot electrons from the corona on the seeds UV photons produced in the accretion disc.
It is the emission we are interested in, since it is strictly associated to the accretion processes onto the SMBH. 
Clear excesses (up to 5$\sigma$) were evident in the $\sim$6-7 keV energy band.
Then, we included, one at a time, three Gaussian lines, with a fixed width of 10 eV.
Given their best-fit energies (see Table \ref{table:xray_analysis_results}), they can be associated with the Fe K$\alpha$ fluorescence emission line at rest-frame 6.39 keV, and the ionised Fe XXV and Fe XXVI fluorescence emission lines at rest-frame 6.7 and 6.97 keV, respectively.
We found no evidence for absorption of the primary continuum emission (N$_H<10^{21}$ cm$^{-2}$). 
Since some residuals were present in the soft part of the analysed band (i.e. at $\sim$2 keV), we included a \emph{mekal} component, needed to model the excess likely produced by the diffuse emission from the high-energy tail of SF.
In the end, our best fit model consists of a power law, three Gaussian emission lines and a thermal component.
The best-fit model ($\chi^2=388.4$ for 321 degrees of freedom) is shown in the top panel of Fig. \ref{fig:xrays}, and the best-fit parameters are presented in Table \ref{table:xray_analysis_results} and are consistent with results in the literature (e.g., \citealt{Emmanoulopoulos13}).
The flux obtained integrating the primary AGN emission --- i.e. the power law --- in the rest-frame 2-10 keV is F$_{\rm 2-10 keV}=(1.22^{+0.01}_{-0.01})\times10^{-11}$ \fluxcgs.\\
Considering the wide energy band (3-27 keV) offered by \nustar, we first fit the continuum emission with a single power law.
In this case, we obtained a poor fit, with significant residual excess in the $\sim$6-7 keV band.
Given the lower spectral resolution provided by \nustar\ ($\sim$400 eV at 6 keV with respect to $\sim$150 eV from \xmm), we were not able to constrain both the centroid and the normalisation of the Gaussian lines needed to model the excess in the $\sim$6-7 keV band.
For this reason, we included the Gaussian functions one at a time, setting the energy in correspondence of the best-fit obtained with \xmm, then we left the normalisation free to vary (as in \citealt{Ursini15}).
The primary emission spectral index is significantly higher ($\Gamma_X=1.81\pm$0.02) than the one observed with \xmm, consistently with the literature \citep{Ursini15}.
This is likely due to the wider energy band available to model the primary emission where there are no significant contributions from other components (e.g., the thermal component below 3 keV).
Part of the primary emission is usually reflected by the surrounding material around the SMBH, resulting in an excess above 10 keV with respect to the continuum.
This reflected component is usually observed in Seyfert 1 galaxies (e.g., \citealt{Perola02}), but has never been observed in NGC 7213. 
We checked for the presence of a reflection component, but the fit was not significantly improved by such inclusion. 
The best-fit is presented in the bottom panel of Fig. \ref{fig:xrays} ($\chi^2=344.2$ for 342 degrees of freedom), while the best-fit parameters are shown in Table \ref{table:xray_analysis_results}.
Integrating the power law over the rest-frame 2-10 keV energy band, we estimated a flux of F$_{\rm 2-10 keV}=(1.62^{+0.02}_{-0.02})\times10^{-11}$ \fluxcgs.\\
Comparing the best-fit results between the two observations, variability in both flux and spectral shape is present.
The observed variability in terms of flux (the flux measured by \xmm\ is $\sim$25\% fainter than the that derived by \nustar\ data) is consistent with what usually observed in AGN, while the different spectral slope ($\Gamma_X=1.64$ and 1.81, see Table \ref{table:xray_analysis_results}) can be due to the different energy band used for the analysis.
In the end, NGC 7213 in the X-rays shows spectral features of a typical low-luminosity Seyfert 1, i.e. $\Gamma_X\sim$1.8, with no evidence for obscuration, and L$_{\rm 2-10 keV}=(1.25\pm0.02)\times10^{42}$ \lumcgs, using the results from the analysis of the \nustar\ observation.
Assuming a bolometric conversion factor of $k_{bol}=9\pm$5 as from \cite{Lusso12}, appropriate for the 2-10 keV luminosity of NGC 7213, we estimate the bolometric luminosity to be L$_{bol}=(1.1\pm0.6)\times10^{43}$ \lumcgs, consistent with previous results from literature (e.g., \citealt{Starling05_I}; \citealt{Emmanoulopoulos13}).
This means that NGC 7213 is accreting at a very low rate, resulting in a rather low fraction of the Eddington luminosity ($\sim$ $9 \times 10^{-4}$, assuming a black hole mass M$_{BH}\sim10^8$ \ms, as from \citealt{WooUrry02}).
This value is relatively low with respect to typical Seyfert 1 galaxies (a few per cent), again stressing the intermediate nature of NGC 7213 between a Seyfert galaxy and a LINER.
\begin{table}[h]
\centering                                      
\begin{tabular}{l c c c}         
\hline\hline                        
Parameter & Value  \\    
\hline   
Parameter & \xmm\ & \nustar\ \\                          
    $\Gamma_{X}$ & 1.64$\pm$0.02 & 1.81$\pm$0.02 \\      
    F$_{\rm 2-10 keV}$ & 1.22$^{+0.01}_{-0.01}$ &  1.62$^{+0.02}_{-0.02}$ \\     
    E$_1$ & 6.40$^{+0.01}_{-0.01}$ \\     
    norm$_1$ & 18.7$^{+1.8}_{-1.4}$ & 16$^{+3}_{-3}$ \\
    E$_2$ & 6.69$^{+0.02}_{-0.03}$ \\     
    norm$_2$ & 5.8$^{+1.1}_{-1.4}$ & 5$^{+3}_{-3}$ \\
    E$_3$ & 6.95$^{+0.06}_{-0.05}$ & 6.95 \\
    norm$_3$ & 3.1$^{+1.7}_{-1.8}$ & 8$^{+3}_{-3}$ \\
    $k$T & 0.4$^{+0.1}_{-0.2}$ &  \\
    norm$_{mekal}$ & 4$^{+2}_{-1}$ &  \\
   \hline                                             
\end{tabular}
\caption{Best-fit parameters from the X-ray spectral analysis using \xmm\ and \nustar\ observations, respectively. From top to bottom: the spectral index ($\Gamma_{X}$), the rest-frame 2-10 keV flux (in units of 10$^{-11}$ \fluxcgs), the energy of the Gaussian emission lines (in units of keV), with their normalisation (in units of $10^{-6}$ photons cm$^{-2}$ s$^{-1}$). In the \xmm\ observation it was necessary to include a thermal component \emph{mekal}, accounting for the excess in the soft band at $\sim$2 keV, likely produced by hot diffuse gas. The plasma temperature of the \emph{mekal}  component is in units of keV, while the normalisation is in units of $10^{-3}$ photons cm$^{2}$ s$^{-1}$.}              
\label{table:xray_analysis_results}
\end{table}
\subsection{ALMA data}
The ALMA observations of NGC 7213 were taken in May 2014 (early science, project: 2012.1.00474.S, PI: N. Nagar) at 230 GHz (Band 6), in configuration C32-5, including 31 12m antennas. 
These observations cover the angular scales in the range 0.5$^{\prime\prime}$-25$^{\prime\prime}$, corresponding to 60 pc - 3 kpc at the redshift of the source, where 0.5$^{\prime\prime}$ is the spatial resolution, while 25$^{\prime\prime}$ is the field-of-view (FoV).
However, the largest angular scale that was recovered with the adopted antenna configuration is 6.2$^{\prime\prime}$, or 750 pc.
The spectral setup consisted of three high-resolution spectral windows with 1920 channels of 976.562 kHz width each, and a low-resolution spectral window of 128 channels of 15.626 MHz width.
Two of the high-resolution spectral windows were centered on the observed-frame frequency of the $^{12}$CO(2-1) and CS(5-4) emission lines at 229.2 GHz and 243.5 GHz, respectively.
The remaining two windows were centred on the sky frequencies at 240.4 GHz and 227.8 GHz, respectively, in order to measure the sub-millimeter continuum emission.\\
The data were calibrated using the ALMA calibration scripts, with CASA version 4.5.3.
J2056-4714 was observed as bandpass calibrator, J2235-4835 as phase calibrator, while Neptune was used as amplitude calibrator, assuming the Butler-JPL-Horizons 2012 model.
From the calibrated data, continuum and line images were obtained using the CASA task \emph{clean}.
We adopted the natural weighting to get the best signal-to-noise ratio.
\subsubsection{The continuum emission}
\label{sec:continuum} 
\begin{figure}[h]
	\includegraphics[width = 0.5\textwidth, keepaspectratio=True]{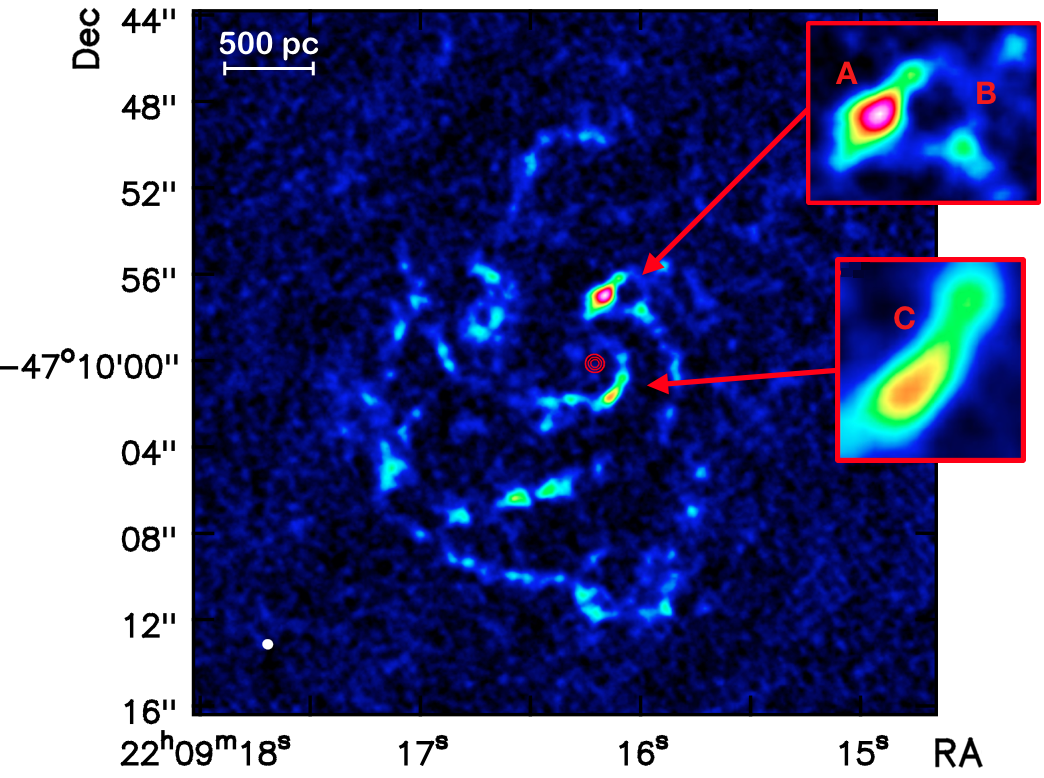}
	\caption{\emph{	ALMA CO(2-1) integrated intensity image with overlaid the continuum emission in red contours (at 5$\sigma$, 10$\sigma$ and 50$\sigma$ level). 
	The white ellipse in the bottom left corner represents the synthesised beam of 0.50$^{\prime\prime}\times0.47^{\prime\prime}$ with a position angle of 73.4 deg.
	 The three interesting regions are magnified in the two boxes: a possible outflow (A) located at the edge of a super-bubble (B) and a second potential outflow observed from the PV diagram analysis (C; see Fig. \ref{fig:pv_diagrams}).}}
	\label{fig:mom0_region}
\end{figure}

In Fig. \ref{fig:mom0_region}, the contour levels of the continuum emission (at 235.1 GHz, or 1.28 mm), obtained from the line-free channels in all the four spectral windows, are presented in red.
The beam size is 0.48$^{\prime\prime}\times0.44^{\prime\prime}$ (with a beam position angle of $\sim$81.1 deg) and a 1$\sigma$ RMS level of 8.9$\times10^{-5}$ Jy beam$^{-1}$.
The emission is clearly produced by a point-like source, of size $\lesssim$60 pc.
Using the \emph{imfit} task from the CASA software, we fitted the map with an elliptical Gaussian profile.
The best-fit centroid is consistent, within the uncertainties, with the NGC 7213 radio position provided by the NASA/IPAC Extragalactic Database (NED)\footnote{https://ned.ipac.caltech.edu/}. 
The \emph{imfit} task provided the continuum flux density, integrating the Gaussian profile, with the corresponding uncertainty: F$_{\nu, \rm cont}=40.1\pm0.1$ mJy.
\subsubsection{The CO(2-1) emission line}
\label{sec:co21} 
\begin{figure}[t]
	\includegraphics[width = 0.5\textwidth, keepaspectratio=True]{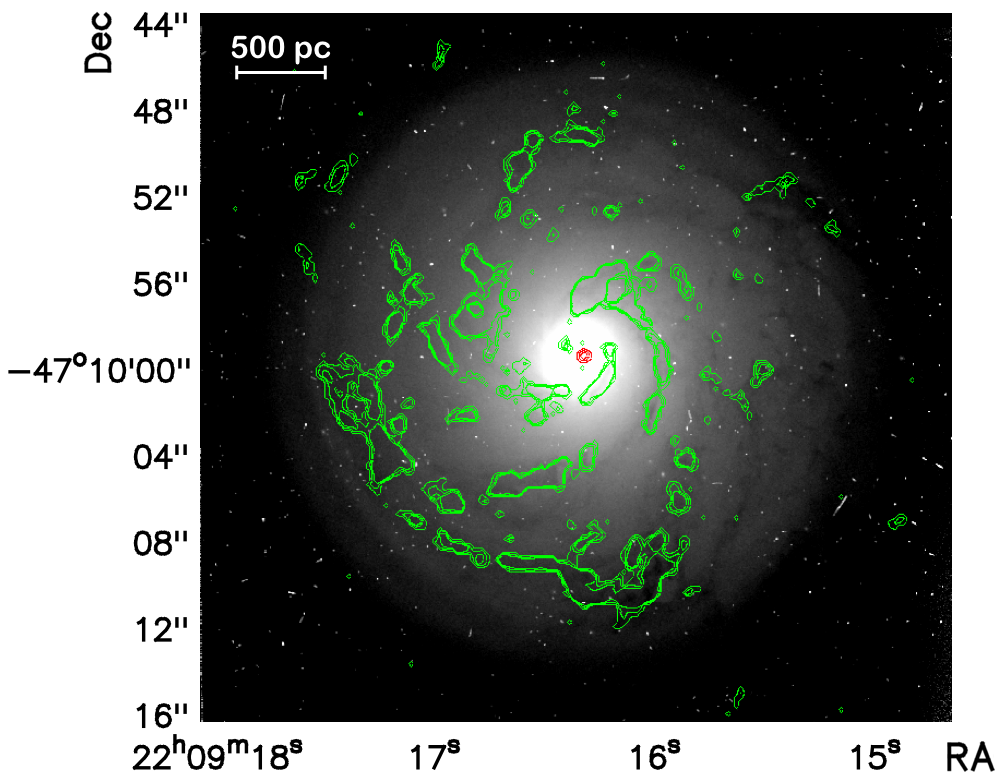}
	\caption{\emph{Contour levels of the continuum (red) and CO(2-1) line emission (green, at 2$\sigma$, 4$\sigma$ and 6$\sigma$ levels) are superimposed on an optical image from the Hubble Space Telescope (\hst; F606W filter). The molecular gas follows the same spiral-like pattern as the optical emission. The continuum is produced by a point-like source.}}
	\label{fig:hst-co}
\end{figure}

The CO(2-1) emission line was extracted from the continuum-subtracted cube of the first spectral window.
We used as reference frequency the CO(2-1) frequency at the redshift of $z=0.0058$ (NED).
We used the \emph{clean} task to iteratively clean the dirty image, selecting a natural weighting scheme of the visibilities.
We binned the cube to increase the signal-to-noise ratio, requiring a spectral resolution of 10 \kms.
The cleaned image of the CO(2-1) emission line has a synthesized beam of 0.50$^{\prime\prime}\times0.47^{\prime\prime}$, with a position angle of 73.4 deg, and an average 1$\sigma$ RMS is 0.1 mJy beam$^{-1}$ per channel.\\
As presented in the integrated intensity map (see Fig. \ref{fig:mom0_region}), the spatial distribution of the CO line flux follows a spiral-like pattern, characterised by a clumpy emission.
This can be explained by the combination of the intrinsic clumpy nature of the emitting medium, with the lack of a more diffuse component, that has most likely been resolved out because of the extended antenna configuration adopted for the interferometric observation.
Using a circular region with a diameter of 25$^{\prime\prime}$ (or $\sim$3 kpc, roughly corresponding to the field-of-view of the instrument), we measured f$_{\rm CO, ALMA}= 112\pm$5 Jy \kms\ as the flux of the CO(2-1) emission line.
The uncertainty on the flux density is the quadratic sum of the two main contributions: the RMS within the same aperture, and the flux calibration uncertainty ($\sim$5\%, as suggested when using Neptune as flux calibrator).
\\
Regarding the morphology, the CO(2-1) emission traces the spiral arms of the galaxy, as can be observed in Fig. \ref{fig:hst-co}, where the contours of the CO emission line are superimposed to an archival \hst\ optical observation (taken with the F606W filter on the WFPC2; \citealt{Malkan98}).
The molecular gas is clearly co-spatial with the spiral arms, while the size of the narrow-line region, estimated from the [O III] line observed with the FR533N filter on the \hst/WFPC2 \citep{Schmitt03}, is less than 100 pc.
This suggests that the CO(2-1) is most likely heated by the stellar activity within the arms rather than the low-luminosity AGN hosted in the center.
This is in agreement with theoretical models (e.g., \citealt{Obreschkow09}; \citealt{Meijerink07}; \citealt{Vallini19}), where the impinging radiation for the low-J transitions like the CO(2-1) mainly comes from the Photo-Dissociation Regions (PDRs; e.g., \citealt{Pozzi17}; \citealt{Mingozzi18}) rather than from the X-ray Dissociation Region (XDR) heated by the central AGN.
Indeed, looking at both Fig. \ref{fig:mom0_region} and Fig. \ref{fig:hst-co}, the lack of CO emission at the location of the ALMA continuum emission and of the peak of the optical emission -- both indicative of the location of the nucleus -- are evident.
\subsection{APEX data}
\label{sec:apex_data}
\begin{figure}[t]
	\includegraphics[width = 0.5\textwidth, keepaspectratio=True]{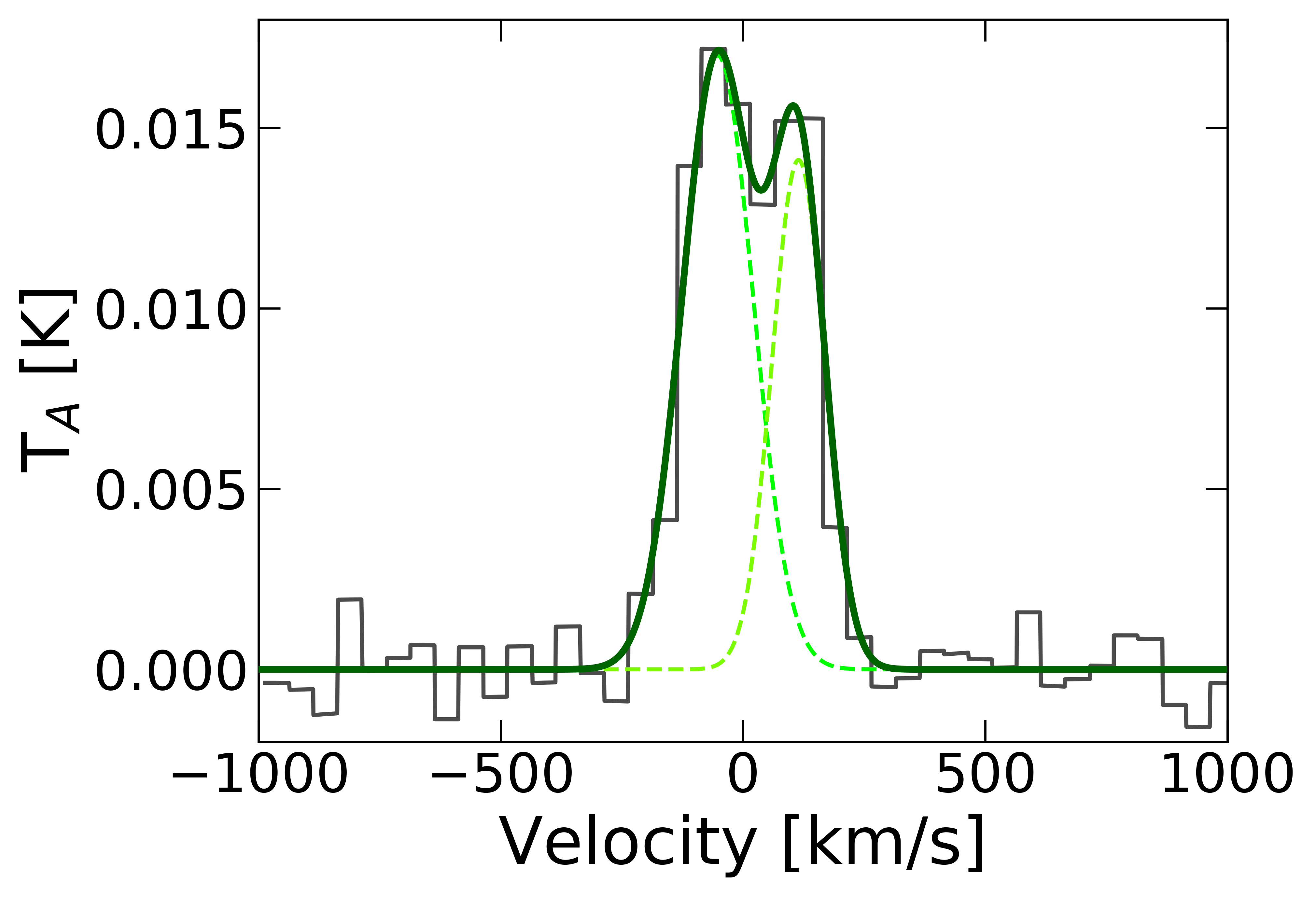}
	\caption{\emph{CO(2-1) emission line observed with the PI230 receiver at APEX.
	On the x-axis the velocity in \kms, on the y-axis the antenna temperature in K; the channel width is $\Delta$v=50 \kms. 
	Two Gaussian functions (in dashed light green lines, while the sum of the two is in darkgreen) are needed to reproduce the double-peak spectrum profile (in black).
	A rotation-dominated kinematics is suggested by the double-peaked line profile.}}
	\label{fig:apex_line}
\end{figure}

The APEX observation of the CO(2-1) emission line (at 229.2 GHz sky frequency) was carried out with the PI230 receiver at the Atacama Pathfinder Experiment (APEX; project 0103.F-9311, PI: F. Salvestrini).
The need for the single-dish observation was motivated by the potential filtering of the CO emission at intermediate and large scales in the interferometric observations.
Indeed, the archival ALMA data were limited by the maximum recoverable scale (MRS; $\sim$6$^{\prime\prime}$, or $\sim$700 pc) of the antenna configuration adopted.
This could result in a significant underestimate of the molecular gas content, traced by the CO emission.
As reported in section \ref{sec:co21}, the clumpy morphology observed in the interferometric observation, along with the lack of a fainter diffuse component, supported this hypothesis.\\
Data reduction was performed using the CLASS program, which is part of the GILDAS\footnote{http://www.iram.fr/IRAMFR/GILDAS/} software.
The CO(2-1) emission line profile is presented in Fig. \ref{fig:apex_line}.
The spectral resolution requested for the APEX observation (50 \kms) is sufficient to observe the double-peak structure of the line profile.
This profile is generally associated with rotation-dominated motion, in agreement with the results from the kinematical study of the ALMA data that is presented in the following section ($\S$ \ref{sec:kinematics}).
The integrated CO(2-1) surface brightness has been obtained performing a fit using a double Gaussian function to the line profile.
The resulting value is $\Sigma_{\rm CO}=9.6\pm$1.4 K \kms.
The uncertainty on the surface brightness is dominated by the calibration uncertainty, which have been conservatively assumed to be 15$\%$ as for similar observations (e.g., \citealt{Csengeri16}; \citealt{Giannetti17}).
To compare this value with the one that we obtained from the ALMA observation, we used the Jy/K conversion factor, which depends on the aperture efficiency of the telescope.
In the configuration adopted for our observation (PI230 detector), with Jupiter as calibrator, the conversion factor is 35$\pm$3\footnote{http://www.apex-telescope.org/telescope/efficiency/}.
Then, f$_{\rm CO, APEX}= 340\pm$60 Jy \kms, i.e. $\sim$3 times f$_{\rm CO, ALMA}$ reported in section \ref{sec:co21}.
This implies that the ALMA interferometric observations only recovered about 30\% of the CO(2-1) flux density measured with APEX.
\subsection{SED decomposition}
\label{sec:sed_fitting}
\begin{figure}[h]
	\includegraphics[width = 0.5\textwidth, keepaspectratio=True]{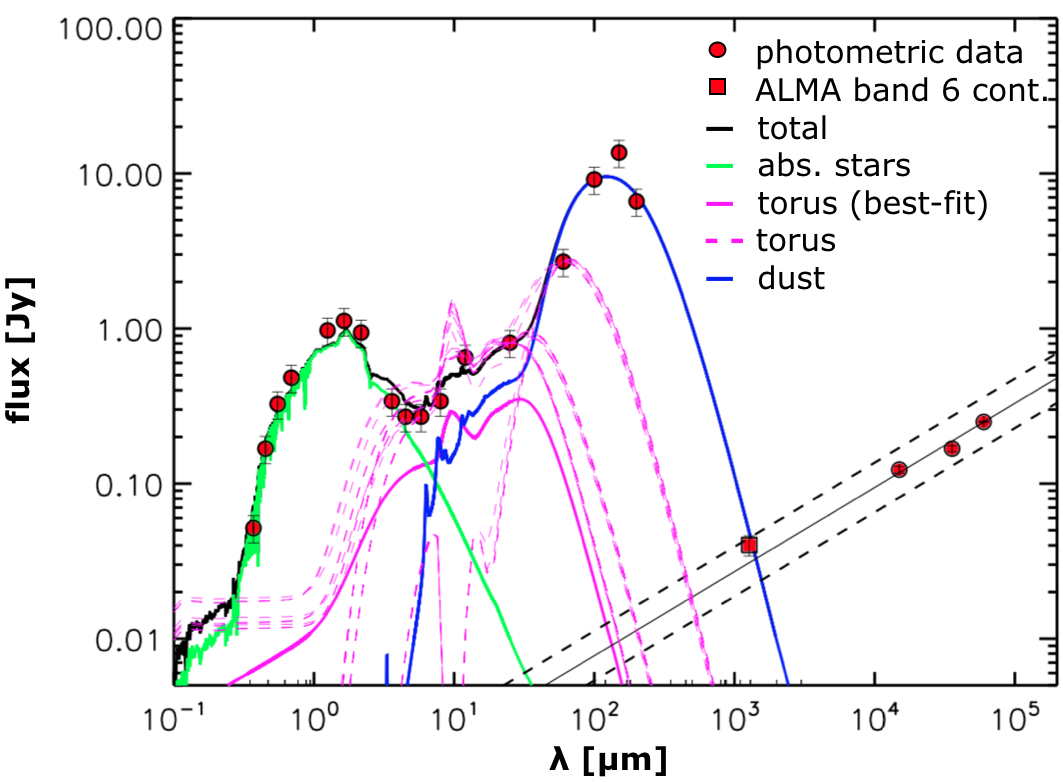}
	\caption{\emph{Decomposed SED of NGC 7213, obtained with the \texttt{SED3FIT} code \citep{Berta13}. Green, pink (continuous line) and blue lines represent the contribution from the extinguished stars, the dusty torus and the emission reprocessed by dust, respectively, while the black solid curve is the sum of all components (total emission).
	The red circles are the photometric measurements, from the optical to the radio frequencies (i.e. the ATCA observations at 5, 8 and 20 GHz, not included in the SED decomposition). 
	As explained in Section \ref{sec:sed_fitting}, the nature of this emission (represented by the red square) is compatible with non thermal emission produced in the nuclear region, i.e. synchrotron emission.
	Even considering extreme models (the pink dashed lines), the dusty torus cannot be the responsible for the observed emission (red square) since, it would be too faint at mm wavelengths (at least two orders of magnitudes fainter).
	Then, we fitted the three radio-frequencies observations by ATCA with a power law, obtaining the best-fit slope presented with the black solid line (the dashed lines correspond to the 1-$\sigma$ levels).}}
	\label{fig:sed_synchrotron}
	\end{figure}

NGC 7213 benefits from a detailed SED decomposition performed by G16, which allowed us to disentangle the relative contributions of AGN and SF activity to the global IR outcome of the source, providing a characterisation of the host galaxy in terms of stellar and dust content (M$_{\star}$ and M$_{dust}$, respectively), and ongoing SF (SFR).
Here we briefly introduce the photometric data collected from the archive, and the SED decomposition procedure adopted by G16.
The homogenised catalogue of total fluxes, from the UV to the FIR, is presented by G16 (see also their Table A.1; the flux densities are corrected for the aperture and magnitude zero-point).
In the case of NGC 7213, the photometric data included in the analysis are: the U, B, V, R bands from \cite{3RC}; the NIR measurements from the catalogue by \cite{2MASS}; the \emph{Spitzer}/IRS spectrum re-binned by \cite{Gruppioni16}, and the photometry by \cite{IRAC} and \cite{IRAS} in the MIR; the FIR photometry by \cite{ISOPHOT}.
The adopted SED-fitting code was \texttt{SED3FIT}\footnote{http://steatreb.altervista.org/alterpages/sed3fit.html} \citep{Berta13}, which reproduces the stellar emission, the emission of the dust heated by the stars and the AGN/torus emission simultaneously.
The code used the library by \cite{BruzualCharlot03} for the stellar contribution, the one by \cite{daCunha08} for the IR dust-emission, and the library of smooth AGN tori by \cite{Fritz06}, updated by \cite{Feltre12}.
In order to limit the degeneracy among the torus parameters, in G16 the AGN configurations of obscured sources were excluded for NGC 7213 (as supported by optical observations of the source and by the X-ray spectral properties presented in Sec. \ref{sec:xray-analysis}).
The best-fit model and the decomposition in the different components is presented in Fig. \ref{fig:sed_synchrotron}.
The host-stellar contribution and the dusty SF dominates over the AGN in the optical bands and in the entire IR band, respectively.
While this could appear to be in contrast with the type 1/broad-line nature of the AGN, it is in agreement with the relatively weak nuclear activity observed in NGC 7213 (revealed also through the X-ray spectral analysis reported Sec. \ref{sec:xray-analysis}).
%
%
\section{Interpreting the CO and continuum sub-mm emission}
\label{sec:interpretation}
%
%
\subsection{The point-like continuum}
\label{sec:continuum_interpretation}
The continuum map at 235.1 GHz (or 1.28 mm) is shown in Fig. \ref{fig:mom0_region}.
Two interpretations are consistent with the point-like nature of the observed continuum: \emph{a)} nuclear synchrotron emission, \emph{b)} thermal dust emission.
The interpretation in point \emph{a)}, i.e. non-thermal emission produced in the very nuclear region, is supported by the compact morphology and the result of the fit presented in Fig. \ref{fig:sed_synchrotron}.
Furthermore, the ALMA continuum emission (represented by a red square, Fig. \ref{fig:sed_synchrotron}) is consistent with the extrapolation of the relation derived from the radio points including uncertainties.
This relation has been obtained by fitting a power-law relation (S$_{\nu}$$\propto$$\lambda^{\alpha}$) to the radio flux densities at 5, 8 and 20 GHz (obtained simultaneously at the Australia Telescope Compact Array, ATCA; \citealt{Murphy10}).
Indeed, the source appears to be point-like (i.e. there are no evidence for jets or large scale structures) at these frequencies.
We obtained a slope $\alpha=0.54\pm$0.03, which is consistent with the slope observed in the case of synchrotron emission.
With the current data, we are not able to exclude the contribution from other mechanisms (e.g., free-free emission; see discussion in \citealt{Ruffa18}).\\
Regarding point \emph{b)}, i.e. the contribution from thermal emission, we refer to the results of the SED de-composition analysis presented in G16 and briefly introduced in Section \ref{sec:sed_fitting}.
Given the point-like nature of the continuum emission observed with ALMA (shown as a red square in Fig. \ref{fig:sed_synchrotron}), we conclude that it cannot be associated with the tail of the FIR bump.
This possibility is rejected since the FIR bump is expected to be produced by a diffuse dust component, which was not detected in the ALMA observation.
Alternatively, it could be associated to the thermal emission from the dusty torus, even if the predicted torus emission at 1.3 mm is significantly lower than the observed flux for the best-fit torus template (see the thick pink line in Fig. \ref{fig:sed_synchrotron}).
To test this hypothesis, we considered also extreme torus configurations (pink dashed lines) to maximise the torus contribution to the FIR emission.
In particular, we considered torus models with high optical depth ($\tau=10$) and with the highest outer-to-inner radius ratio (R$_{max}$/R$_{min}=300$).
Since the dust sublimation radius is usually assumed as the inner radius, with a size of $\sim$pc for a typical AGN luminosity as in the case of NGC 7213 (e.g., \citealt{Fritz06}), this corresponds to an outer radius of $\sim$300 pc.
It is important to notice that the ``extension" of a typical torus in an intermediate-luminous AGN is below 10 pc, as observed with ALMA in local active galaxies (e.g., \citealt{GarciaBurillo14}).
Having said this, also considering these extremely extended torus models, we were not able to reproduce a significant fraction of the continuum emission observed with ALMA.
Then we excluded the thermal emission from dust to be the major contribution to the observed continuum emission at 1.3 mm.
%
%
\subsection{The molecular gas kinematics}
\label{sec:kinematics}
\begin{figure}[h]
  	\includegraphics[width = 0.5\textwidth, keepaspectratio=True]{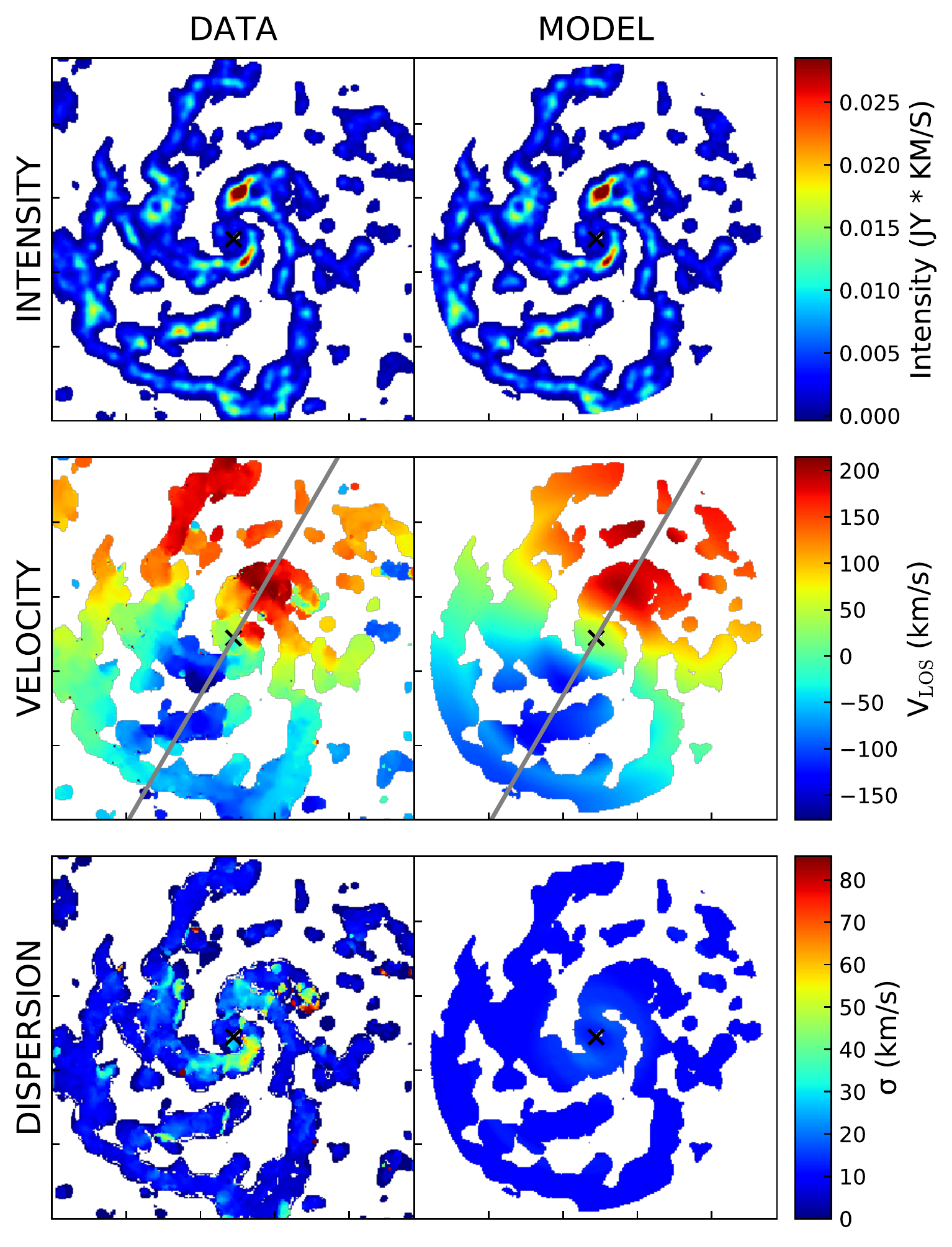}
	\caption{From top to bottom, \emph{comparison between the flux, velocity and velocity dispersion maps obtained from the data (left column) and the best-fit model produced by the \barolo\ code (right column). Both data and model have been convolved with an elliptical Gaussian having a dimension twice the size of the beam of the observation. This smoothing procedure helps the \barolo\ in the fitting procedure, especially in the case of clumpy emission as the one analysed here. The grey lines in the central panels represent the direction of the position angle; the same slit was used to extract the position-velocity (PV) diagram on the major axis, as presented in the top panel of Fig. \ref{fig:pv_diagrams}.}}
	\label{fig:data_model_comparison}
\end{figure}

\begin{figure}[h]
	\includegraphics[width = 0.5\textwidth, keepaspectratio=True]{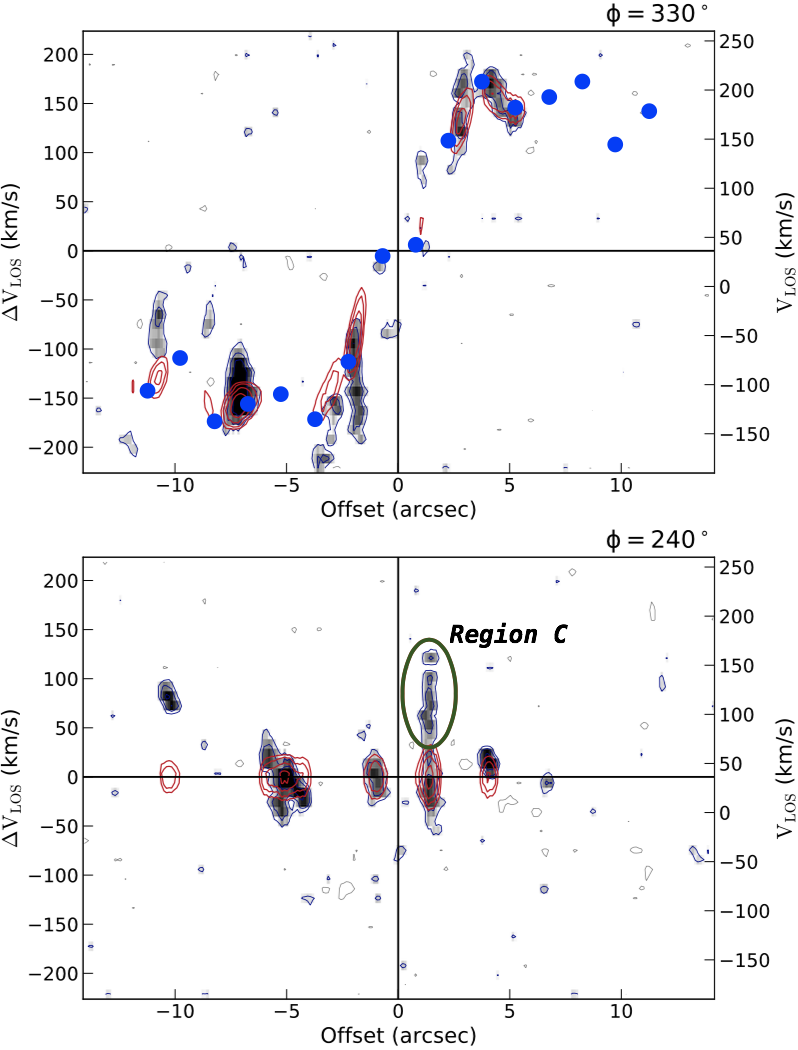}
	\caption{\emph{Position-Velocity diagrams on the major (top panel) and minor (bottom panel) axis, respectively. 
	On the x-axis, the angular distance from the centre of the source along the direction fixed by the angle $\phi$ is reported.
	On the right y-axis, the velocity along the line-of-sight (in \kms) is shown as it appears in the input datacube. On the left, the velocity along the line-of-sight (in \kms) is shown, once the systemic velocity (v$_{sys}=36\pm$10 \kms) has been subtracted.
	The data are represented in grey scales with blue contours, while the the best-fit model is identified by red contours.
	In the upper panel, the blue circles represent the projected best-fit rotational velocity at different radial distances from the center, associated with each ring of the disc.
	In the bottom panel, the signature of the potential outflow associated with \emph{region C} (bordered in green; see section \ref{sec:regions}) is clearly visible as an excess of CO emission over 100 \kms\ with an offset of $\sim$1$^{\prime\prime}$ from the centre of the rotation.}}
	\label{fig:pv_diagrams}
\end{figure}

The spatial resolution provided by ALMA observations allows us to perform a detailed study of the kinematics of the molecular gas, tracing it from the large scales (e.g., the rotating galactic disc) to the very central regions, where the accretion onto the central SMBH takes place.
This kind of study has been successfully performed on a large numbers of local Seyfert galaxies (e.g, \citealt{GarciaBurillo14}), and also in galaxies from the same parent sample (e.g., \citealt{Sabatini18}).
Following a similar approach to that adopted by \cite{Sabatini18}, we used the \barolo\ (3D-Based Analysis of Rotating Object via Line Observations) software \citep{3DBAROLO} to model the kinematics of the molecular gas, as traced by the CO(2-1) emission line.
\barolo\ is a code specifically developed to fit 3D tilted-rings models on two main assumptions: \emph{i}) the material which is responsible for the observed emission has to be contained in a thin disc; \emph{ii}) the kinematic has to be dominated by rotation.
The first assumption is generally accepted for local S0 galaxies.
Regarding assumption \emph{ii}), the ALMA observation of NGC 7213 revealed a rotation-dominated pattern, clearly visible from the velocity map (moment-1 map) of the CO(2-1) emission line (see the central panel of Fig. \ref{fig:data_model_comparison}).
We fixed the kinematic centre to the centroid of the continuum emission, whose profile have been fitted (using the \emph{imfit} task from CASA) using an elliptical Gaussian.
This is based on the assumption that the nucleus is the centre of the rotation and is responsible for the continuum emission (see Section \ref{sec:continuum_interpretation}).
To reduce the model degeneracies, we set the disc geometry fixing some parameters.
From the data, we set the position angle of the major axis to PA$^{\rm major}=330^\circ$ from North to West. 
We also fixed the inclination of the disc with respect to the line of sight to $i=30^\circ$, as estimated in literature works (e.g., \citealt{StorchiBergmann96}; \citealt{Lin18}).
The central velocity of the ALMA data cube has been chosen to be that of the CO(2-1) sky frequency, i.e., 229.2 GHz, but we left the systemic velocity (v$_{\rm sys}$) free to vary in order to account for potential inaccuracy relative to the adopted sky frequency.
Given the tradeoff of ``holes'' and clumpy emission, we prefer the pixel-by-pixel normalisation (i.e., \emph{local}) over the \emph{azimuthal} one (i.e., the azimuthally averaged flux in each ring) in order to account for the non-axial symmetry of the emission, i.e. regions with anomalous gas distribution, which could affect the global fit.
For the same reason, we allow \barolo\ to perform a smoothing of the input data cube by a factor of 2 of the original beam of the observation (i.e. the data have been convolved with an elliptical Gaussian having dimension twice the size of the original beam), and to cut the smoothed cube at a signal-to-noise ratio of 4.
We set the maximum number of radii to 8, separated by 1.5$^{\prime\prime}$ or $\sim$180 pc, excluding the outermost part of the field of view, where \barolo\ was not able to fit the model to the faint clumpy emission.
To summarise, the free parameters are: the circular velocity, the systemic velocity and the velocity dispersion.\\
At first, we did not include any potential radial velocity components in order to reduce the number of free parameters.
Once the best-fit model was obtained, i.e. the residuals (data minus model) showed no significant evidence for rotational motion, we tested for the presence of a radial velocity component (v$\_{rad}$).
We run the code with this additional free parameter, but its best-fit value was consistent with zero.
By assuming different values of the PA (PA$=320^{\circ}-340^{\circ}$) and inclination ($i=20^{\circ}-40^{\circ}$), one at a time, we tested the goodness of the fiducial values of the PA and inclination angle.
Since the residual clumps resulted to be not sensitive to the choices of the assumptions, we preferred the best-fit model with PA$^{\rm major}=330^\circ$ and $i=30^\circ$.
The results from the kinematical analysis are presented in Fig. \ref{fig:data_model_comparison}, where the comparison between the model and the data in terms of intensity, velocity and velocity dispersion maps (zero-th, first and second moments, respectively) is shown.
The model and data (smoothed by a factor 2) are in excellent agreement.
The kinematics of the molecular gas, traced by the CO(2-1) emission line, is clearly dominated by purely circular rotational motion around the nucleus.
This is the first successful attempt to model the kinematics of the molecular gas in NGC 7213, thanks to the software \barolo\, which is able to handle the available data, despite the sparse information which limited previous attempts to model the kinematics (e.g., \citealt{Ramakrishnan19}).
As clearly visible in Fig. \ref{fig:pv_diagrams}, we found that the data cube central velocity was offset by an additional $\delta$V$_{\rm sys}=36\pm$10 \kms.
This means that the systemic velocity of the source with respect to our rest frame has to be V$_{\rm sys}=1716\pm$10 \kms, which is a mean value between the systemic velocity obtained from the study of the stellar kinematics by \cite{Schnorrmuller14}, and the results by \cite{Ramakrishnan19}.\\
Looking at the lower panels, the velocity dispersion in the best-fit model is $\sim10-15$ \kms, values which are expected from a rotation-dominated disc.
The highest values of the velocity dispersion ($v_{disp} \sim$70 \kms) are not reproduced by the best-fit model.
The high-dispersion residual clumps resulted to be insensitive to our assumptions (e.g. PA, $i$, v$_{rad}$), therefore we suggested that the residuals are associated with non rotational motion, as discussed in the following section.
%
%
\subsection{Complex structures in the CO emission}
\label{sec:regions}
\begin{figure}[h]
  	\includegraphics[width = 0.5\textwidth, keepaspectratio=True]{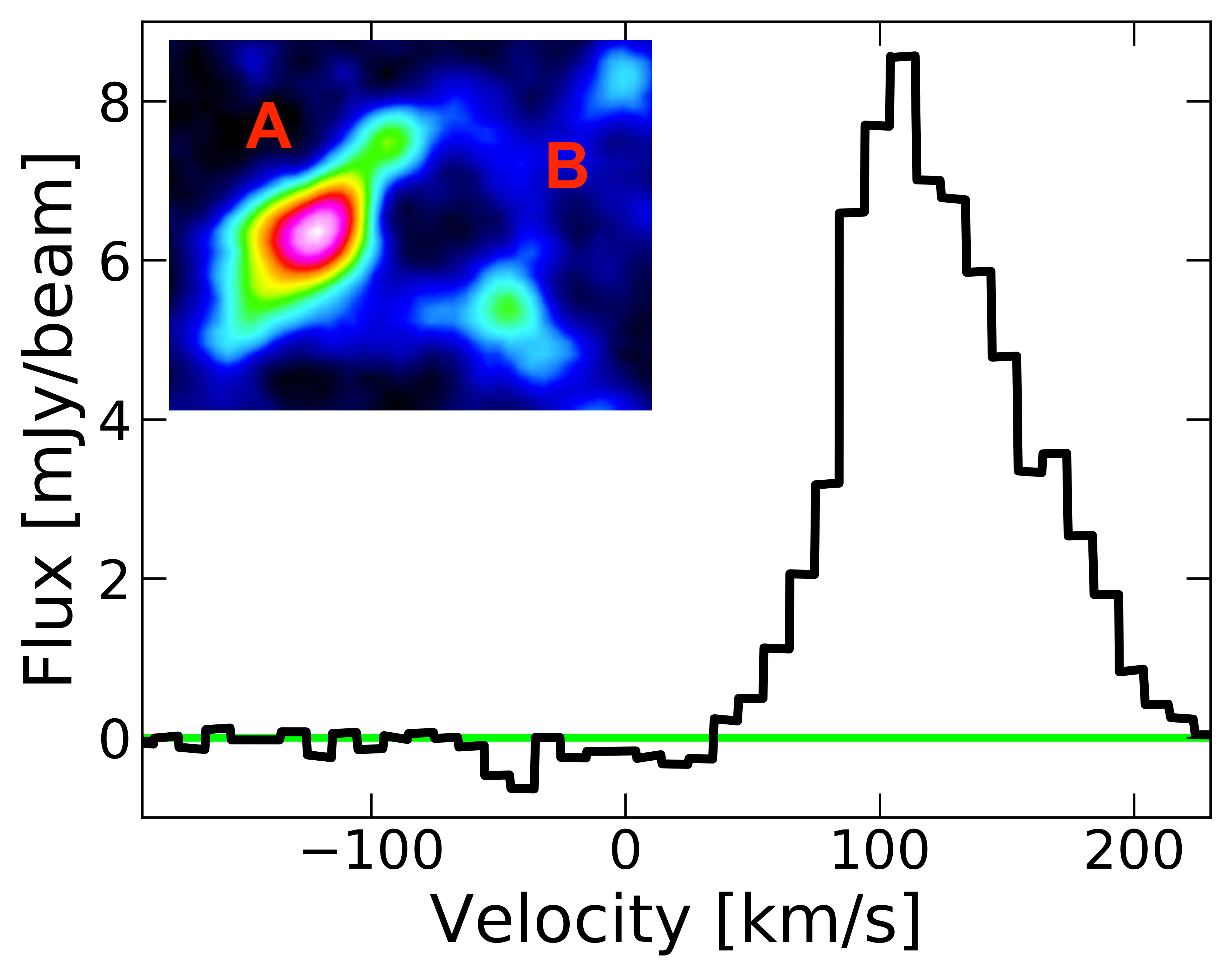}
	\caption{\emph{The asymmetric line profile observed in correspondence with the potential outflow in \emph{Region A}.
	This emission is likely produce by shocked material.}}
	\label{fig:regA}
\end{figure}

The results of the modelling of the gas kinematics presented in section \ref{sec:kinematics} pointed out the presence of few regions where the kinematics is not strictly associated to rotational motion.
To investigate the nature of these regions, we used both the position-velocity diagrams (PV diagrams), shown in Fig. \ref{fig:pv_diagrams}, and the moment maps (intensity, velocity and velocity dispersion maps) from Fig. \ref{fig:data_model_comparison}.
In particular, we focused on three regions, that we named \emph{A}, \emph{B} and \emph{C} (see Fig. \ref{fig:mom0_region}), that were interpreted as follows:\\ 
\emph{Region A}: an extended emission (of the order of 1.5$^{\prime\prime}$x2.5$^{\prime\prime}$), showing an asymmetric line profile (see Fig. \ref{fig:regA}).\\
\emph{Region B}: a possible super-bubble, i.e. a nearly circular void of diameter 0.7" or 90 pc, surrounded by likely shocked material. It is located 3.3" (400 pc) NE from the nucleus.\\
\emph{Region C}: a likely outflow located 1.4" (150 pc) SE from the nucleus along the minor axis.\\
Since \emph{Region A} is located at the edge of the circular void identified as \emph{Region B}, the first can be interpreted as the emission produced by the shock front impacting on the surrounding medium.
However, further observations able to recover the CO emission at all scales with good spatial resolutions are needed to validate this hypothesis. 
The potential outflow in \emph{Region C} was identified from the position-velocity (PV) diagram along the minor axis, presented in the bottom panel of Fig. \ref{fig:pv_diagrams}.
This emission (\emph{Region C}), located along one of the spiral arms and observed over 100 km/s (corresponding to at least 10 channels), is clearly not consistent with what is expected from a rotation-dominated disc.
The interpretation as emission from outflowing gas is supported by the presence, at the same location, of a peak in velocity dispersion, associated with a region of a size 1.5$^{\prime\prime}$x1$^{\prime\prime}$, or 180 pc $\times$120 pc (see the dispersion map in Fig. \ref{fig:data_model_comparison}).
From the residual of the kinematical modelling, we measured the peak velocity ($v_{max}$), the size ($R_{OF}$), and the flux of the emission associated with the outflow.
The flux was used to derive the outflow mass ($M_{OF}$), following the procedure extensively described in Section \ref{sec:gas-mass}.
The molecular gas outflow mass was calculated by assuming an \aco=1.1 \uaco, as for the gas mass from the APEX observation, which is a value similar to the one used in literature (e.g., \citealt{Cicone14}; \citealt{Fiore17}).
Then, we computed the mass-outflow rate ($\dot{M}_{OF}$) of \emph{Region C}, assuming a spherical geometry (i.e., $\dot{M}_{OF}=3v_{max}M_{OF}/R_{OF}$).
We compared this value ($\dot{M}_{OF}\sim0.03\pm0.02$ M$_{\odot}$ yr$^{-1}$) with the one we expected from the $\dot{M}_{OF}-L_{AGN}$ relation (e.g., \citealt{Cicone14}; \citealt{Fiore17}), predicting for \emph{Region C} a mass loss $\dot{M}_{OF}$$\sim$7-15 M$_{\odot}$ yr$^{-1}$.
This prediction is at least two orders of magnitudes larger than what we measured, thus suggesting that is unlikely to be purely AGN driven.
Furthermore, the outflow is located relatively distant from the low-luminous nucleus of NGC 7213, thus suggesting a dominant contribution from a stellar-like driven mechanism.\\
The same calculation has been performed in the case of the potential outflow in \emph{Region A}, suggesting a stellar-related mechanism powering the molecular wind, given the measured $\dot{M}_{OF}\sim0.05\pm$0.03 M$_{\odot}$ yr$^{-1}$.
%
\subsection{The molecular gas mass}
\label{sec:gas-mass}
The cold molecular gas mass (M$_{\rm gas}$) can be estimated from the luminosity of the CO(1-0), given \aco, the CO-to-H2 mass conversion factor (M$_{\rm gas}=$\aco L$^{\prime}_{\rm CO(1-0)}$; see \citealt{SolomonVandenBout05}).
This relation is widely used in the literature, but \emph{a}), we need to extrapolate the CO(1-0) luminosity from our measurements of the CO(2-1) line; \emph{b}), we need to assume an appropriate value for the \aco.
Regarding point \emph{a}), while at high transition numbers (J>3) the CO spectral energy distribution (COSLED) strongly depends on the excitation mechanism (i.e. SFR, AGN, shocks; \citealt{Meijerink07}; \citealt{Pozzi17}; \citealt{Mingozzi18}) and the ISM physical properties (i.e. density, geometry, see \citealt{HollenbachTielens99}; \citealt{Vallini19}), at lower-J the COSLED shape is rather uniform, tracing the cold and diffuse phase (e.g., \citealt{NarayananKrumholz14}), with a typical CO(2-1)-to-CO(1-0) flux ratio $\sim$3 (e.g., \citealt{Papadopoulos12}).
We therefore assumed $f_{\rm CO(2-1)}$/$f_{\rm CO(1-0)}=3$, and we applied this factor to the CO(2-1) emission line flux observed with APEX.
We did not take into account the CO(2-1) line flux observed with ALMA because, as explained in section \ref{sec:apex_data}, it represents only a fraction of the flux observed with APEX.
This difference is likely due to the filtering out of the large scale emission in the interferometric observation, hence considering the CO flux measured with ALMA could result in a significant underestimate of M$_{\rm gas}$.\\
Then, we have calculated L$^{\prime}_{\rm CO(1-0)}$ following \cite{CarilliWalter13}:\\
L$^{\prime}_{line}=3.25\times10^7\times S_{line}\Delta v\frac{D_{L}^2}{(1+z)\nu_{rest}^2}$ \uLco,\\
where $S_{line}\Delta v$ is the flux of the emission line in Jy \kms, assuming for the CO(1-0) line width the same as measured for the CO(2-1), $D_L$ is the luminosity distance in Mpc, $\nu_{rest}$ the rest-frame frequency of the line in GHz.
We obtained L$^{\prime}_{\rm CO(1-0), APEX}=(1.8\pm0.3)\times10^8$\uLco.
To check whether the CO emission recovered by APEX observation is representative of the expected molecular gas emission from NGC 7213, we use the relation log(L$^{\prime}_{\rm CO(1-0)}$)=(0.73$\pm$0.03)log(L$_{\rm IR}$)+(1.24$\pm$0.04) (e.g., \citealt{CarilliWalter13}), which relates the CO and the IR luminosity (L$_{\rm IR}$).
A negligible contribution from the AGN in the FIR is assumed, as actually observed (see also Fig. \ref{fig:sed_synchrotron}).
Using the SF-related IR luminosity provided by G16 (L$_{\rm IR}=(1.0\pm0.3)\times10^{10}$\ls), we obtained L$^{\prime}_{\rm CO(1-0)}=(4\pm2)\times10^8$\uLco.
The CO luminosity obtained with APEX is roughly a factor $\sim$2 smaller than the one extrapolated from the IR emission; however, the two values are consistent within the uncertainties.
This suggests that the CO luminosity collected within the APEX aperture (25$^{\prime\prime}$, or $\sim$3 kpc) is almost representative of the CO emission from the entire galaxy, hence of the entire molecular gas content.
Therefore, for the estimate of M$_{\rm gas}$ presented in this section, we used the CO(1-0) luminosity from APEX.\\
The choice of an appropriate \aco\ factor (point \emph{b}) depends on the ISM conditions.
\aco\ has a strong dependance on the metallicity (metal-poor galaxies show \aco\ up to a factor of 5-10 higher than Milky Way-like galaxies, see Fig. 9 from \citealt{Bolatto13}) and a relatively less significant dependance on the compactness/starburstness of the sources (see Fig. 12 from \citealt{Bolatto13}).
Since we cannot derive the metallicity for NGC 7213 from the optical spectrum, we assume it to be solar-like.
For active and luminous IR galaxies, the generally adopted \aco\ values are in the range $\sim$0.3-2.5 \uaco (see also \citealt{DownesSolomon98}; \citealt{Papadopoulos12}). 
For NGC 7213, we assumed \aco=1.1 \uaco, which is used for the nuclear regions of metal-rich galaxies (e.g., \citealt{Sandstrom13}; \citealt{Rosario18}).
Thus we compared our result with literature works on local active galaxies showing similar properties to NGC 7213 (e.g., \citealt{Pozzi17}; \citealt{Rosario18}).
Based on the above assumptions, the molecular gas mass derived from the flux estimate obtained with APEX is M$_{\rm gas}=(2.0\pm0.3)\times10^8$ \ms.\\
Given M$_{\rm gas}$, we estimated the depletion time, as t$_{\rm depl}=$M$_{\rm gas}$/SFR.
Since the SFR provided by G16 (SFR$=1.0\pm0.1$ \ms\ yr$^{-1}$) is referred to the entire galaxy, while the APEX aperture covers $\sim25^{\prime\prime}$ (or $\sim$3 kpc), we needed to scale down the SFR.
This is necessary since the two measurements were obtained with different apertures, and because the SF surface brightness in NGC 7213 appeared to be extended beyond the central region \citep{Diamond-Stanic12}. 
We considered the \emph{Herschel}/PACS observation at 70 $\mu$m (PSF FWHM$\sim$5.6$^{\prime\prime}$), where the contribution from the central AGN should be less important to the global IR outcome with respect to the relative nuclear contribution at shorter wavelengths, as suggested by the result of the SED decomposition analysis presented in Fig. \ref{fig:sed_synchrotron} and from the low-luminosity nature of NGC 7213 presented in this work.
The extended IR emission at 70 $\mu$m is almost entirely produced by the SF activity, hence it can be used as a proxy of the SFR surface brightness.
We found the ratio of the flux within the entire galaxy (F$_{tot}$) and the APEX aperture (25$^{\prime\prime}$; F$_{25^{\prime\prime}}$)) being F$_{tot}$/F$_{25^{\prime\prime}}$=2.0$\pm$0.3.
Eventually, assuming the scaled SFR, we measured a depletion time to be t$_{\rm depl}$=0.4$\pm$0.1 Gyr.
This values is consistent with what is observed in the local Universe in objects with similar properties in terms of M$_{\star}$ and SFR ($0.1<t_{\rm depl}<$few Gyr; e.g., \citealt{Rosario18}), but larger than what is observed at high redshift (0.01$<t_{\rm depl}<$0.1 Gyrs; e.g., \citealt{Brusa18}; \citealt{Kakkad17}; \citealt{Talia18}).
The observed difference in t$_{\rm depl}$ between high and low redshift samples is most likely due to the stronger SF and AGN activity at the cosmic noon (e.g., \citealt{MadauDickinson14}), resulting in shorter time scales for the gas consumption.
%
%
\section{Discussion and conclusions}
\label{sec:conclusions}
In this work we have presented a multi-wavelength approach to the study of NGC 7213, a low-luminosity AGN with a wealth of multi-wavelength observations.
The source was selected from the sample presented by G16, on the basis of the quality of the available archival observations in the X-rays and at mm wavelengths.
The multi-band information helps us in drawing a more complete picture of the physical processes in this object, the different phases of the ISM in the host galaxy and of the role of the AGN.
To this aim, we performed a spectral analysis of X-ray archival observations to study the accretion-related emission in terms of power and spectral shape, and we combined the high-sensitivity and high-spectral resolution from ALMA, crucial to trace the molecular gas kinematics down to sub-kpc scales, with the spatially integrated information provided by APEX, to estimate the molecular gas content.
The main results of this work can be summarised as follows:
\begin{itemize}
\item Our re-analysis of archival \xmm\ and \nustar\ observations allows to proper characterise the central engine in terms of spectral shape and power over the 2$-$27 keV energy band.
The results of the X-ray spectral analysis (i.e. $\Gamma_X=1.81\pm$0.02 and F$_{\rm 2-10 keV}=1.62^{+0.02}_{-0.02}$ \fluxcgs, derived from \nustar\ observation due to the wide energy band covered; see Table \ref{table:xray_analysis_results}) support the presence of an unobscured AGN in the center of NGC 7213, in agreement with the classification as Seyfert 1 galaxy, based on the optical broad line features.
However, the relatively low luminosity of the source in the X-rays (L$_{\rm 2-10 keV}\sim1\times10^{42}$ \lumcgs) suggests a low accretion rate, way below the Eddington limit.
The energetics related to the nuclear activity of NGC 7213 places the source in an intermediate stage between a typical Seyfert galaxy and a LINER.

\item Using \barolo\ on the ALMA data of the CO(2-1) emission, we obtained the first model of the molecular gas kinematics in the central regions of NGC 7213.
The best-fit model well reproduces the velocity fields, which is dominated by a rotational pattern.
From the residuals we found no evidence for non rotationally-dominated motion in the central region (i.e. $\lesssim$60 pc from the nucleus).
This means that the SMBH hosted in NGC 7213 is not able to affect significantly the motion of the molecular gas traced by the CO(2-1) at the scales recovered in the available observation, since there is no evidence for nuclear inflows or molecular gas streaming feeding the AGN.
\item The study of the CO emission line data cube showed some evidence for two potential outflows, located within 500 pc from the nucleus.
\emph{Region A} is located at the edge of a circular void, likely a super-bubble. 
The evidence for \emph{Region C} came from the PV-diagram analysis and is located along one of the spiral arms.
Given the sizes and the location of both, they are more likely powered by stellar activity, rather than by the AGN, but better data are needed to confirm this hypothesis.
\item The continuum emission at 235.1 GHz (1.28 mm) is produced by a point-like source.
Based on an SED analysis, we concluded that the most reasonable interpretation is the continuum being produced through synchrotron radiation in the nuclear region, in agreement with the extrapolation of the ATCA observations at longer wavelengths.
\item The molecular gas mass of NGC 7213 is M$_{\rm gas}=(2.0\pm0.3)\times10^8$ \ms, obtained converting the CO luminosity observed with APEX.
We underline how the ALMA observation would have underestimated the gas mass by a factor $\sim$3, given the filtering out of the large scale emission in interferometric observations.
We estimate a depletion time of t$_{\rm depl}=0.4\pm$0.1 Gyr, which is consistent with what is observed in local moderately luminous Seyfert galaxies (e.g., \citealt{Rosario18}).
This suggests a negligible, if any, impact of the AGN on the host-galaxy SF activity.
\end{itemize} 
The proposed approach allowed us to combine the available multi-wavelength information to obtain a coherent picture of the source in terms of both AGN activity and host-galaxy ISM properties.
In the case of NGC 7213, the accretion-related emission from the AGN is rather weak, unable to impact significantly the molecular gas content and distribution of the galaxy, hence to influence the SF activity, as suggested by the depletion time.
Given the results of our study, NGC 7213 can be classified as a LLAGN, showing an intermediate nature between a Seyfert 1 galaxy and a LINER.\\
In a future work we plan to apply the same multi-wavelength approach to all the objects of the G16 sample with similar multi-wavelength observations in order to assess, in a statistical manner, the impact of the AGN on the ISM of their hosts.\\
%
%
%
%
%
%
\begin{acknowledgements}
We thank Dr. F. Fraternali and Dr. di Teodoro for the valuable discussions about the use of \barolo\ and the interpretation of the results of the modelling of the molecular gas kinematics. 
We also thank G. Sabatini for his availability in helping to use the \barolo\ code.
Based on observations collected at the European Southern Observatory under ESO programme 0103.F-9311(A).
The time granted was used to obtained data for the target of this work.
The research leading to these results has received funding from the European Union\'s Horizon 2020 research and innovation programme under grant agreement No 730562 [RadioNet].
This paper makes use of the following ALMA data: ADS\/JAO.ALMA\#2012.1.00474.S. ALMA is a partnership of ESO (representing its member states), NSF (USA) and NINS (Japan), together with NRC (Canada), MOST and ASIAA (Taiwan), and KASI (Republic of Korea), in cooperation with the Republic of Chile.
The Joint ALMA Observatory is operated by ESO, AUI\/NRAO and NAOJ.
The co-author C. Vignali acknowledges financial support from the Italian Space Agency (ASI) under the contracts ASI-INAF I/037/12/0 and ASI-INAF n.2017-14-H.0.
\end{acknowledgements}
%
%
\bibliographystyle{aa}
\bibliography{ngc7213_biblio.bib}

\end{document}